# Single-Pixel Imaging in Space and Time with Optically-Modulated Free Electrons

Andrea Konečná[1,2], Enzo Rotunno[3], Vincenzo Grillo[3], F. Javier García de Abajo[1,4,*], and Giovanni Maria Vanacore[5,*]

1. ICFO-Institut de Ciencies Fotoniques, The Barcelona Institute of Science and Technology, 08860 Castelldefels (Barcelona), Spain
2. Central European Institute of Technology, Brno University of Technology, 612 00 Brno, Czech Republic
3. Centro S3, Istituto di Nanoscienze-CNR, 41125 Modena, Italy
4. ICREA-Institució Catalana de Recerca i Estudis Avançats, Passeig Lluís Companys 23, 08010 Barcelona, Spain
5. Laboratory of Ultrafast Microscopy for Nanoscale Dynamics (LUMiNaD), Department of Materials Science, University of Milano-Bicocca, Via Cozzi 55, 20121 Milano (Italy).

* To whom correspondence should be addressed:
javier.garciadeabajo@nanophotonics.es, giovanni.vanacore@unimib.it

**Abstract**
Single-pixel imaging, originally developed in light optics, facilitates fast three-dimensional sample reconstruction, as well as probing with light wavelengths undetectable by conventional multi-pixel detectors. However, the spatial resolution of optics-based single-pixel microscopy is limited by diffraction to hundreds of nanometers. Here, we propose an implementation of single-pixel imaging relying on attainable modifications of currently available ultrafast electron microscopes in which optically-modulated electrons are used instead of photons to achieve sub-nanometer spatially- and temporally-resolved single-pixel imaging. We simulate electron beam profiles generated by interaction with the optical field produced by an externally programable spatial light modulator and demonstrate the feasibility of the method by showing that the sample image and its temporal evolution can be reconstructed using realistic imperfect illumination patterns. Electron single-pixel imaging holds strong potential for application in low-dose probing of beam-sensitive biological and molecular samples, including rapid screening during in-situ experiments.

**INTRODUCTION**
Single-pixel imaging (SPI) is a key application of structured-wave illumination. This method, which has been recently developed in the context of optical imaging, relies on the interrogation of a certain object using a number of spatially-modulated illumination patterns while synchronously measuring the total intensity of the scattered light captured by a single-pixel detector [1,2,3,4]. Key elements in this method are (1) a spatial light modulator (SLM), which provides the spatial encoding of the illumination patterns that is necessary for image reconstruction, and (2) the inherent 'sparsity' of typical real-space images, such that the bulk of the information is only contained in a limited number of pixels, and consequently, compressed sensing (CS) can be used [5,6,7,8]. CS uses prior knowledge of sparsity in the coefficient domain, making the reconstruction of the image possible by using a smaller number of measurements. Specifically, O(Klog(N)) measurements are typically needed, if the information is K-sparse and has N pixels.

The idea behind SPI is to perform a number of sequential measurements with specific illumination patterns expressed on a sufficiently complete basis that can be either incoherent (random patterns) or spatially-correlated (such as Hadamard or Fourier bases) with the object to be imaged. The ensemble of *M* measurements, identified by the vector $\chi$, is then correlated to the image ***T*** (sample transmission function) with a number of pixels $N_{pix}$ (in which one usually has *M* << $N_{pix}$) through the *M*×$N_{pix}$ measurement matrix ***H***, which contains the employed SLM patterns, such that $\chi = HT$. An image reconstruction algorithm is then used to retrieve a reconstructed image $T^*$.

In optical microscopy, the SPI technique has a well-established tradition and its unique measurement scheme has demonstrated far superior performance with respect to conventional imaging. This is because the illumination patterns used for sampling can be custom-tailored to maximize the amount of information acquired during the measurement, whereas in conventional imaging information gathering is bound to stochastic processes. Different aspects of this idea have been the topic of recent relevant literature in the field of SPI. In particular, several groups have demonstrated that the ordering of Hadamard patterns, for instance, is of primary importance to maximize the effectiveness of CS algorithms. Different ordering based on the significance of the patterns (i.e. different a-priori knowledge) have been proposed, such as, to mention a few, the "Russian Dolls" ordering [9], the "cake cutting" ordering [10], the "Origami pattern" ordering [11] and an ordering based on the total variation of the Hadamard basis [12]. This concept can be pushed to its ultimate limit when deep learning is used to gather a-priori information and identify the best set of illumination patterns [13]. In this way, it has been demonstrated that, in a limiting scenario in which an object must be identified within a restricted pool of choices, the task can be accomplished without even needing to reconstruct the image [14], but just after a single SPI measurement. Incidentally, compressed sensing approaches have recently been used in TEM for encoding temporal dynamics in electron imaging with 10 kHz frame rate (100 µs resolution) [15].

In SPI, the number of illumination patterns required for high-quality imaging increases proportionally with the total number of pixels. However, CS methods and, more recently, deep learning (DL) approaches have been considered to substantially reduce the number of measurements necessary for the reconstruction of an image with respect to the total number of unknown pixels. This is an extremely interesting aspect for electron microscopy, since it would entail lower noise, faster response time, and lower radiation dose with respect to conventional imaging. DL approaches, which have already demonstrated superior performances with respect to CS in terms of speed and sampling ratio, can be organized into three categories: (1) Improving the quality of reconstructed images [16,17,18,19]; (2) identifying the best illumination strategy by exploiting the features learned during training [13,14]; and (3) reconstructing the target image directly from the measured signals [19,20,21,22,23]. Also, a reduction in the sampling rate well below the Nyquist limit (down to 6%) has been demonstrated using DL.

Such advantages would be particularly appealing in the context of electron imaging of nano-objects in their biological and/or chemical natural environment, for which the minimization of the electron dose is critical [24,25] to avoid sample damage. Initial attempts have been made using MeV electrons with beam profiles controlled by laser image projection on a photocathode [26].

This method is, however, incompatible with the sub-nanometer resolution achieved in transmission electron microscopes (TEMs) through electron collimation stages. Sub-nanometer resolution for SPI thus requires patterning of high-quality coherent beams. In TEMs, SPI has never been proposed and adopted before, mainly due to the lack of fast, versatile, and reliable electron modulators that would be able to generate the required rapidly changing structured electron patterns.

Here, we propose to implement SPI in a TEM by illuminating the sample using structured electron beams created by a Photonic free-ELectron Modulator (here referred as PELM). The PELM is based on properly synthesized localized electromagnetic fields that are able to create an efficient electron modulation for programable time/energy and space/momentum control of electron beams. Our approach adopts optical field patterns to imprint on the phase and amplitude profile of the electron wave function – an externally-controlled well-defined modulation varying both in time and space while the electron pulse crosses the light field. The PELM concept relies on the ability to modulate electrons with optical fields [27,28,29,30,31] down to attosecond timescales [32,33,34,35,36] and along its transverse coordinates [37,38,39,40]. In essence, we overcome the problem of designing and fabricating complicated electron-optics elements by resorting to shaping light beams, which has been proven a much easier task to perform, while in addition it enables fast temporal modulation. Indeed, a critical advantage of our approach with respect to existing methods lies in the possibility of achieving an unprecedented ultrafast switching and an extreme flexibility of electron manipulation, which can also open new quantum microscopy applications [41,42].

A suitable platform for generating the required light field configurations is represented by a light-opaque, yet electron-transparent thin film on which an externally controlled optical pattern is projected from a SLM. The SLM provides an out-of-plane electric field, $E_z(x,y)$, with a customized transverse configuration that embodies the required laterally-changing phase and amplitude profiles. In such a configuration, the spatial pattern imprinted on the incident light field by the SLM is directly transferred onto the transverse profile of the electron wavepacket, as recently shown both theoretically [43,44] and experimentally [45,46]. Different portions of the electron wave profile experience a different phase modulation as dictated by the optical pattern. We can thus obtain an externally-programable electron beam with a laterally-changing encoded modulation. Moreover, the ability to modulate the electron phase and amplitude has the potential to overcome Poisson noise [47,48], which is a key aspect that renders the SPI method not only feasible but also advantageous in terms of low-dose imaging.

A synchronized intensity measurement followed by a CS or DL reconstruction could then be used to retrieve the sample image. Of course, the possibility to use CS or DL algorithms strictly relies on the amount of *a priori* information known about the object under investigation [7]. This is particularly relevant for ESPI, which can benefit from such *a priori* information, especially in terms of optimal discrimination, more than conventional imaging. In fact, standard TEM imaging is generally object-independent and any *a priori* information is applied only after acquisition to interpret the image, something that can be understood as a de-noising procedure. Instead, SPI allows one to optimize the acquisition strategy even before starting the experiment and, thus, holds a direct advantage when using the appropriate pattern basis (see Supplementary Information for a direct example).

In Fig. 1a-c, we present the different single-pixel schemes that can be implemented in an electron microscope for 2D spatial imaging (Fig. 1a), 1D spatial imaging (Fig. 1b), and 1D temporal reconstruction (Fig. 1c). Specifically, 2D spatial imaging involves the use of a basis of modulation patterns changing in both transverse directions *x* and *y* (for instance, a Hadamard basis) for full 2D image reconstruction. Instead, 1D spatial imaging involves the use of modulation patterns changing only along one direction (such as a properly chosen Fourier basis) coupled to temporal multiplexing of the electron beam on the detector, which should enable a simpler and faster 1D image reconstruction.

The third scenario of temporal reconstruction is conceptually novel. Importantly, the 1D single-pixel reconstruction algorithm works for any dependent variable of the system phase space. This implies that, by choosing a well-defined basis of temporally-changing modulation functions, such as a series of monochromatic periodic harmonics, it would be possible to reconstruct the time dynamics of a sample. The nature of the method would also allow us to reconstruct the dynamical evolution on a temporal scale much smaller than the electron pulse duration because the resolution depends only on the different frequency components of the basis and not on the length of the electron wavepacket. In principle, it could be even be implemented with a continuous electron beam.

**RESULTS**

### *Principles of Single-Pixel Imaging*

Single-pixel imaging relies on pre-shaped illumination intensity patterns $H^m(\mathbf{R_S})$ that are transmitted through a sample described by a spatially-dependent amplitude transmission function $T(\mathbf{R_S})$ – defining the sample image –, such that the intensity collected at the detector associated with the $m^{\text{th}}$ illumination pattern is

$$\chi^m = \int d^2\mathbf{R_S} T(\mathbf{R_S}) H^m(\mathbf{R_S}), \tag{1}$$

where we integrate over the sample plane and $\chi^m$ are the elements of the measurement vector. The target is to reconstruct the sample transmission function

$$T(\mathbf{R_S}) = \sum_m t^m H^m(\mathbf{R_S}) \tag{2}$$

in terms of coefficients $t^m$. Now we assume that the illumination patterns in general yield

$$\int d^2\mathbf{R_S} H^m(\mathbf{R_S}) H^{m'}(\mathbf{R_S}) = S^{mm'}, \tag{3}$$

where $S^{mm'}$ are real-valued coefficients. Now, by substituting Eq. (2) in Eq. (1), we retrieve

$$\int d^2\mathbf{R_S} \sum_m t^m H^m(\mathbf{R_S}) H^{m'}(\mathbf{R_S}) = \chi^{m'}. \tag{4}$$

Notice that if the illumination patterns form an orthonormal basis, we immediately recover $t^m = \chi^m$ (i.e., the intensities recorded at the detector can directly serve as the expansion coefficients). However, in the general case, where Eq. (3) holds, the expansion coefficients are

$$t^m = \sum_{m'} \chi^{m'} (S^{-1})^{mm'}. \tag{5}$$

By substituting the coefficients back in Eq. (2), we find the general formula

$$T(\mathbf{R_S}) = \sum_m \sum_{m'} \chi^{m'} (S^{-1})^{mm'} H^m(\mathbf{R_S}) \tag{6}$$

for the reconstruction of the sample transmission function. It is worth noting that, besides our current choice, many different orthogonalization algorithms have been implemented in the literature (see for instance Ref. [49]), which can also be used in combination with our ESPI scheme.

### *Single-Pixel Imaging in a TEM via a Photonic Electron Modulator*

We now proceed to analytically describe the scheme utilized to implement the SPI method in an electron microscope. This is shown in Fig. 2, where the sample illumination is performed using structured electron beams created via light-induced manipulation. Efficient and versatile phase and intensity modulation of a free electron can be achieved using a PELM device. In our configuration, the spatial pattern imprinted on the incident light field by a programable SLM is transferred on the transverse profile of the electron wavepacket by electron-light interaction [45]. This is generally dubbed as the Photon-Induced Near-Field Electron Microscopy (PINEM) effect [27,28,50], although in our configuration we actually exploit the breaking of translational symmetry induced by a thin film (inverse transition radiation) – as described in detail in Ref. [35,51,52] – rather than a confined near field induced by a nanoscale structure. The shaped electron wavepacket is then propagated through the TEM column towards the sample.

The electron-light interaction under consideration admits a simple theoretical description [35,43]: starting with an electron wave function $\psi_0$ incident on the PELM, after interaction with the light field, the electron wave function is inelastically scattered into quantized components of amplitude

$$\begin{aligned}\psi_\ell^m(\mathbf{R}_{PELM}) &= \psi_0(\mathbf{R}_{PELM}) J_\ell(2|\beta^m(\mathbf{R}_{PELM})|)\exp(i\ell\, arg\{-\beta^m(\mathbf{R}_{PELM})\}) \\ &= \psi_0(\mathbf{R}_{PELM})\mathcal{F}\{\beta^m(\mathbf{R}_{PELM})\},\end{aligned} \quad (7)$$

corresponding to electrons that have gained ($\ell > 0$) or lost ($\ell < 0$) $\ell$ quanta of photon energy $\hbar\omega$. Here, $\mathcal{F}\{...\}$ represents the PINEM operator, which depends on the imprinted variation of the transverse profile, governed by the coupling coefficient

$$\beta^m(\mathbf{R}) = \frac{e}{\hbar\omega}\int dz\, E_z^m(\mathbf{R})\exp(-i\omega z/v), \quad (8)$$

where $\hbar$ is the reduced Planck constant, $e$ is the elementary charge, and the light illumination is further characterized by the electric field $\mathbf{E}^m$ (see Supplementary Information for a detailed calculation of $\beta^m$ in a metallic thin film). We assume that beam electrons have velocity $\mathbf{v} \parallel \hat{z}$. Due to the inelastic nature of the PINEM interaction, post-interaction electrons gain or lose different numbers of quanta, associated with kinetic energy changes $\ell\hbar\omega$. In addition, the corresponding contributions to the wave function in Eq. (7) have different spatial distributions of amplitude and phase. For our purpose, it would be beneficial to place a simple energy filter after the PELM, selecting, for example, the $\ell = 1$ component only (i.e., electrons gaining one photon energy quantum).

The energy filter needs to efficiently separate a given sideband of the electron energy distribution from the rest of the spectrum. The higher the filter efficiency, the larger the contrast in the modulation pattern, also resulting in a more reduced noise in the final image. However, a relatively modest reduction should be sufficient, as we estimate that ~34% of the electron signal can be placed in the first (gain or loss) sideband. In addition, as we are interested in intensity patterns, the first gain or loss sidebands both deliver the same pattern, and thus, 68% of the electrons are contributing by simultaneously filtering both bands. As a possible improvement, light patterns could be also engineered to eventually remove the need for energy filtering. These possibilities are in fact enabled by properly tuning the light field intensity and, thus, the resulting modulation of the electron beam and its energy distribution [35].

A practical approach towards the design of the structured beam sample illumination is to define a suitable $\psi_\ell^m$ and thus also $\beta^m$, study the propagation of the wave function to the sample plane, and then find optimal settings for the aperture size, beam energy, and focal distance in such a way that $H^m(\mathbf{R}_S)$ mimics the optical illumination pattern. For ESPI, we thus impose the inelastically scattered electron wave function, $\psi_\ell^m$, to be equal to the target pattern, $\psi_T$, defined within the chosen basis ($\psi_\ell^m = \psi_T$). Once this is defined, we can retrieve the coupling coefficient, $\beta^m$, and, therefore, the light field, $E_z^m$, to be implemented on the SLM by applying an inverse PINEM transformation, $\mathcal{F}^{-1}\{...\}$, to the target pattern $\psi_T$ (see Fig. 2 and also Supplementary Fig. S1 for a Hadamard basis, and Supplementary Fig. S2 for a Fourier basis).

Particularly important is to demonstrate the feasibility of the method also under realistic, non-ideal conditions. We do this by applying a momentum cutoff ($\omega_0/nc$, where $n = 1, 2, 3$) on the retrieved light field defined by a momentum-dependent point spread function (PSF) to take into account the finite illumination wavelength and range of angles. This produces the actual light field, $E_z^m|_{\text{actual}}$, from which we can calculate the actual coupling coefficient, $\beta^m|_{\text{actual}}$. By applying the PINEM transformation, $\mathcal{F}\{...\}$, we can in turn find the actual target pattern, $\psi_T|_{\text{actual}}$.

The sequence of operations is defined in Eq. (9) below and visually shown in Figures 2, S1, and S2:

$$\psi_\ell^m = \psi_T \to \beta^m = \mathcal{F}^{-1}\{\psi_T\} \to \beta^m|_{\text{actual}} = \beta^m * \text{PSF}_n \to \psi_T|_{\text{actual}} = \mathcal{F}\{\beta^m|_{\text{actual}}\}. \qquad (9)$$

In order to maximize the efficiency of the electron amplitude and phase modulation, it is beneficial to place the PELM onto a plane along the microscope column where the beam is extended to diameters much larger than the wavelength of the optical illumination. In such a scenario, we can achieve the desired detail in the variation of the transverse wave function profile. However, we then have to rely on electron lenses to focus the beam on the sample.

The focusing action together with the free propagation of the electron wave function between the PELM and the sample planes is described, within the paraxial approximation, as [43]

$$\psi_S^m(\mathbf{R}_S, z_S) \approx \frac{-i\xi}{2\pi} \exp[iq_0(z_S - z_{PELM})] \cdot$$
$$\cdot \exp(-i\xi R_S^2/2) \int d^2\mathbf{R}_{PELM} \, \psi_T|_{\text{actual}}(\mathbf{R}_{PELM}) P(\mathbf{R}_{PELM}) \cdot \qquad (10)$$
$$\cdot \exp\left[iq_0 R_{PELM}^2/2\left(\frac{1}{z_S - z_{PELM}} - \frac{1}{f}\right)\right] \exp[-i\xi(x_{PELM}x_S + y_{PELM}y_S)],$$

where we have defined $\xi = q_0/(z_S - z_{PELM})$ with $q_0$ the electron wave vector that varies with acceleration voltage, and the coordinates $\mathbf{R}_S = (x_S, y_S)$ evolving in the sample $z = z_S$ plane. In addition, $P(\mathbf{R}_{PELM})$ is a transmission (pupil) function, which becomes 1 if the electron beam passes through an effective aperture placed in the PELM plane and 0 otherwise. We have also replaced the focusing action of all subsequent lenses by a single aberration-free thin lens with a focal distance $f$ placed virtually just after the PELM. The illumination intensity at the sample resulting from Eq. (10) is

$$I^m(\mathbf{R}_S) = |\psi_S^m(\mathbf{R}_S, z_S)|^2 \propto \frac{\xi^2}{4\pi^2} |\int d^2\mathbf{R}_{PELM} J_1(2|\beta^m(\mathbf{R}_{PELM})|)\exp(i \arg\{-\beta^m(\mathbf{R}_{PELM})\})$$

$$\times \exp\left[iq_0 R_{PELM}^2/2\left(\frac{1}{z_S - z_{PELM}} - \frac{1}{f}\right)\right]\exp[-i\xi(x_{PELM}x_S + y_{PELM}y_S)]|^2. \tag{11}$$

In Fig. 2b and Supplementary Figs. S1 and S2, we show the realistic sample patterns obtained for a Hadamard pattern and a Fourier pattern, chosen as examples when using the following parameters: 200 keV electrons, lens focal distance $f = 1$ mm, $z_{PELM} = 0$, $z_S = 1.0008$ mm (defocus of 0.8 μm), and PELM area of 10x10 μm².

It is important to mention that the ESPI method here proposed is based on electron intensity modulation, rather than phase modulation. Therefore, there are no stringent constraints or requirements on the transverse coherence of the electron beam for the method to work properly. This is what makes this technique readily available in many different experimental configurations where, for instance, one would favor electron current density over coherence to increase the signal-to-noise-ratio of the measurements. Of course, if the transverse coherence of the electron beam is commensurate with the spatial scale at the PELM plane in which a significant phase change of the interaction strength $\beta^m$ takes place, then phase modulation effects could be visible. Under such conditions, the method could take advantage of the possibility to imprint also a phase modulation – besides an amplitude modulation – on the electron transverse profile. This aspect would not only largely increase the number of patterns forming the basis used for the reconstruction, but it could also potentially allow us to image phase objects via the ESPI method in analogy to optical SPI [53].

An efficient reconstruction can be achieved with a binary illumination using the Hadamard basis, where $N$ sample pixels (e.g., discrete $\mathbf{R}_S$ points) can be reconstructed with $N$ patterns [54]. However, because the Hadamard basis adopts +1 and -1 values to ensure orthogonality, in our case non-orthonormality issues might arise from the fact that we are working with intensity patterns that are never negative. This aspect, together with the imperfect illumination under realistic, non-ideal conditions (see Fig. 2), implies that the actual sample patterns no longer represent an orthonormal basis, and therefore, the reconstructed sample transmission function, $T(\mathbf{R}_S)$, has to the corrected as described in Eqs. (5) and (6) via the overlap matrix $S^{mm'}$. The latter and its inverse are shown in Supplementary Fig. S3 for the Hadamard basis. Another option for a basis is to use Fourier-like intensity patterns (Fourier basis), which are defined as

$$H(\mathbf{R}_S, \mathbf{K}, \varphi) = a + b\cos(\mathbf{K} \cdot \mathbf{R}_S + \varphi), \tag{12}$$

where $\mathbf{K}$ are spatial frequencies, $a$ and $b$ are constants, and $\varphi$ is a phase.

## DISCUSSION

### *Image Reconstruction Using Hadamard and Fourier Bases*

We show next several examples of image reconstruction using different bases. We consider a Siemens star and a ghost image. The former is a binary {0,1} image with sharp transitions, whereas the latter presents small features, is asymmetric, and shows a gradual intensity variation from 0 to 1. This allows us to test in full the capabilities of the method.

In Fig. 3a, we plot the ideal and reconstructed Siemens star and ghost image considering different cutoffs for a Hadamard basis. Clearly, the reconstructions reproduce all the main features of the original images, although we also encounter some noise and even a few negative values, which should not appear. The latter is due to ill-conditioned matrix inversion that we need to use for the reconstruction to compensate for the non-orthonormality of the involved patterns. In Fig. 3b, we plot the results of image reconstruction using a Fourier basis. Although for the Siemens-star reconstruction with the Fourier basis is performing similarly as with the Hadamard basis, for the ghost image it is clear that the Fourier basis with the same number of patterns (64 x 64) yields artifacts: a faint *mirror-reflected* ghost is superimposing on the actual one. The reconstruction with the Fourier basis becomes considerably better when taking into account a phase offset, so that the Fourier pattern would no longer be symmetric with respect to the origin. In Fig. 3c, we have considered an offset of $\varphi = \pi/4$ for the corresponding reconstructed sample images. As a result, the reconstructed ghost image no longer exhibits the faint *mirror-reflected* artifact that was visible in Fig. 3b.

Based on the results of Fig. 3, we performed additional quantitative analysis on the images in order to compare the different reconstruction algorithms and bases. We have extracted the peak signal-to-noise ratio (PSNR) for both the Siemens star and the Ghost images for the two bases and three different cutoff frequencies used. From these calculations, we conclude that the reconstruction with a Fourier basis provides values of the PSNR about 10% better than the Hadamard basis for all cutoffs. This is probably due to the fact the Hadamard basis is composed of binary patterns, which are extremely sensitive to distortions caused by diffractive effects during electron propagation, whereas such effects are mitigated for Fourier patterns, which are characterized by gradual, smooth variations. The better quality of the images reconstructed via Fourier patterns directly implies a better image resolution. This is visible in Fig. 3d, where we show the effect of the reconstruction on the spatial shape of a particularly sharp feature of the Siemens star. As expected, we observe an increasing broadening when smaller cutoff frequencies are considered. The estimated spatial resolution (for a 10:90 fit of the error function) varies from 0.29 nm at a cutoff of $\omega_0/c$ to 1.43 nm at a cutoff of $\omega_0/3c$ for the Hadamard-reconstructed images, whereas the Fourier basis provides slightly better values ranging from 0.25 nm at a cutoff of $\omega_0/c$ to 1.01 nm at a cutoff of $\omega_0/3c$.

It is important to mention that the ESPI method that we propose here is intended for imaging amplitude objects. In fact, in TEM a huge amount of information resides in amplitude contrast mechanisms, such as mass-thickness contrast, Z-contrast, Bright-field and Dark-Field imaging, as well as Electron Energy-Loss Spectroscopy (EELS) and Energy Dispersive X-Ray Spectroscopy (EDX). In a standard TEM, single-pixel detectors are in fact already present. This is for instance the case of the High Angle Annular Dark Field (HAADF) detector used for performing Z-contrast imaging in STEM mode, which can also provide an experimental verification of the proposed configurations. Besides their use as single-pixel detectors, STEM detectors are also able to gather signals in different angular regimes. Such capability is generally used to access simultaneously more information about the sample (typically chemical information). In the SPI context, we can anticipate a more complex partition of the detector – exploiting its angular detection capability – bridging the gap with other techniques such as Integrated Differential Phase Contrast (iDPC) or ptychography.

### *Temporal Electron Single Pixel Imaging*

As a final aspect, we present a possible implementation of the 1D temporal ESPI reconstruction scheme. The basic idea is to be able to reconstruct the dynamic behavior of a sample – for instance, its dielectric response to an optically-induced electronic excitation – using a sequence of temporally-modulated electron pulses with varying periodicity. In Fig. 4a, we show the schematics of the experiment, where a sequence of long light pulses with varying periods $T_j$ couple to the electron pulse via inverse transition radiation as mediated by the metallic plate. The longitudinally-modulated electron pulse then interacts with the sample in its excited state and, for each period $T_j$, a signal $I_j$ is measured. In terms of the single-pixel formalism, this means that we are choosing a one-dimensional Fourier-like basis for the evolution of the incident electron current as a function of time with respect to the pumping time:

$$H^m(t) = e^{-\frac{\left(t - \frac{t_{\max} + t_{\min}}{2}\right)^2}{2\sigma^2}} \sin^2[\pi m \, t/(t_{\max} - t_{\min})], \tag{13}$$

where $t_{\min}$ and $t_{\max}$ determine the boundaries of the sampling time interval and $\sigma^2$ is the variance of the envelope of the probing electron wave function.

As discussed in detail in the Supplementary Information, we have simulated the dynamics of a system comprising three states (A, B, and C) according to the diagram in Fig. 4b. At time zero, the system is taken to be pumped to an excited state A, from which it decays in a cascade fashion to B and then to C. The time evolution of populations of the three states within our model system is governed by three rate equations. In Fig. 4b, we show the results of a temporal Fourier reconstruction using the basis functions defined by Eq. (13) in an analogous way to the spatial domain and, again, taking into account the non-orthogonality of the illumination basis. We demonstrate that already with 20 basis functions the gross features of the temporal response of the system are retrieved.

It is important to note that the temporal resolution of the measurement no longer depends on the duration of the electron and light pulses, but only on the frequency bandwidth of the light field used for electron modulation. This aspect is extremely interesting because it opens the possibility of using continuous electron and light beams, provided that an efficient electron-light coupling is achieved [55,56,57,58,59]. A possible technological implementation of such scheme can be realized by using an optical parametric amplifier (OPA) coupled to a difference frequency generator (DFG). Such a configuration would provide light fields with periods in the 0.8 fs – 50 fs range, making our approach invaluable to investigate sample dynamics with a temporal resolution that is far below that of state-of-the-art ultrafast electron microscopy, and equally combined with the atomic spatial resolution provided by electron beams.

**Acknowledgements**
This work is part of the SMART-electron project that has received funding from the European Union's Horizon 2020 Research and Innovation Programme under grant agreement No 964591. We also acknowledge partial support by the European Research Council (Advanced Grant No. 789104-eNANO) and the Spanish MINECO (Severo Ochoa CEX2019-000910-S).

**Competing Interests**
All authors declare that they have no competing interests.

**Author contributions**
G.M.V, V.G. and F.J.G.d.A. conceived the idea. A.K. and F.J.G.d.A. performed the PINEM calculations. A.K., E.R. and V.G. performed the SPI calculations. G.M.V., E.R., and A.K. performed data analysis. All authors participated in interpreting the results, data discussion and manuscript preparation.

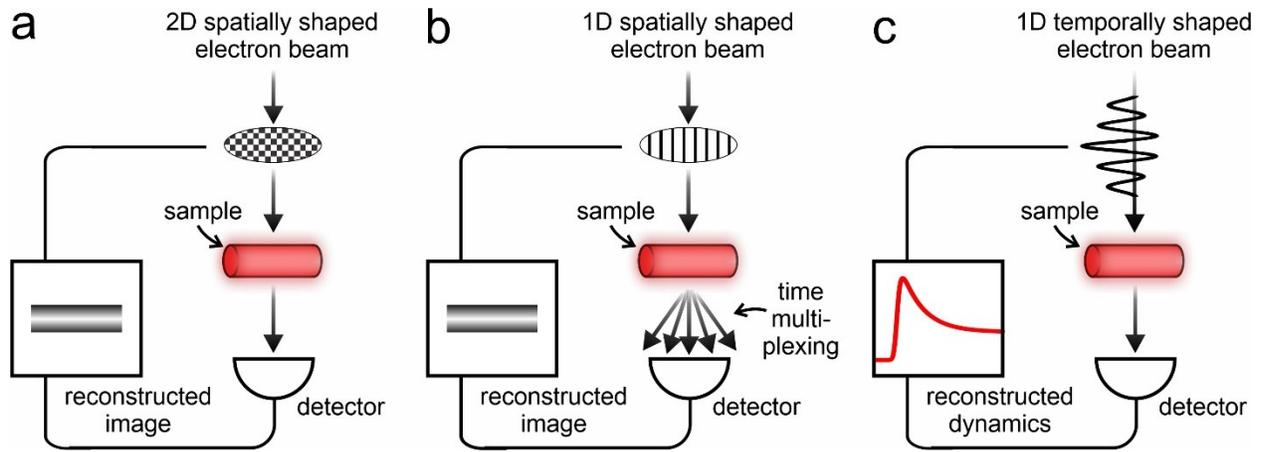

**Figure 1. Single Pixel Imaging with electrons.** Schematic representation of different single-pixel schemes that are amenable to implementation in a transmission electron microscope for 2D spatial imaging (*panel a*), 1D spatial imaging (*panel b*), and 1D temporal reconstruction (*panel c*).

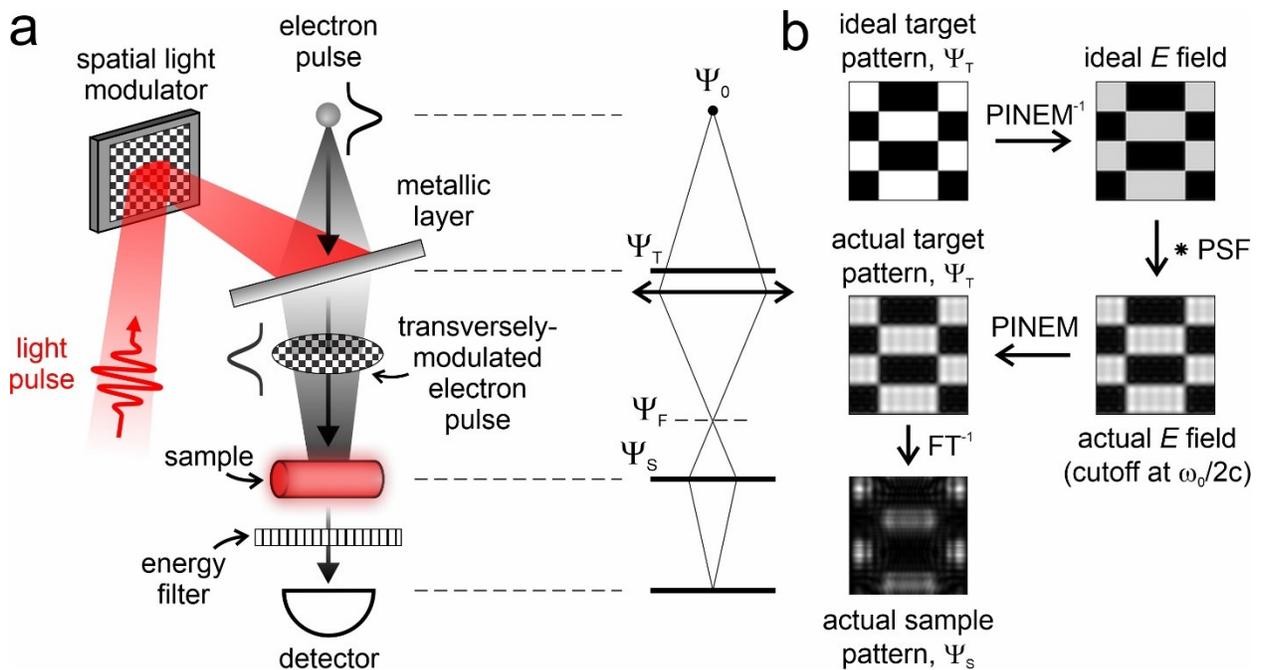

**Figure 2. Electron Single Pixel Imaging (ESPI) via light-mediated electron modulation.** *Panel a*: Schematic representation of the experimental layout considered for the single-pixel imaging method, implemented by using structured electron beams that are in turn created via light-based manipulation. In our configuration, the spatial pattern imprinted on the incident light field by a programable spatial light modulator is transferred on the transverse profile of the electron wavepacket by electron-light interaction. *Panel b*: Sequence of operations used to calculate the transverse distribution of the electron beam arriving on the sample either when starting from an ideal target pattern or when considering realistic non-ideal conditions. We take a pattern from a Hadamard basis for this example.

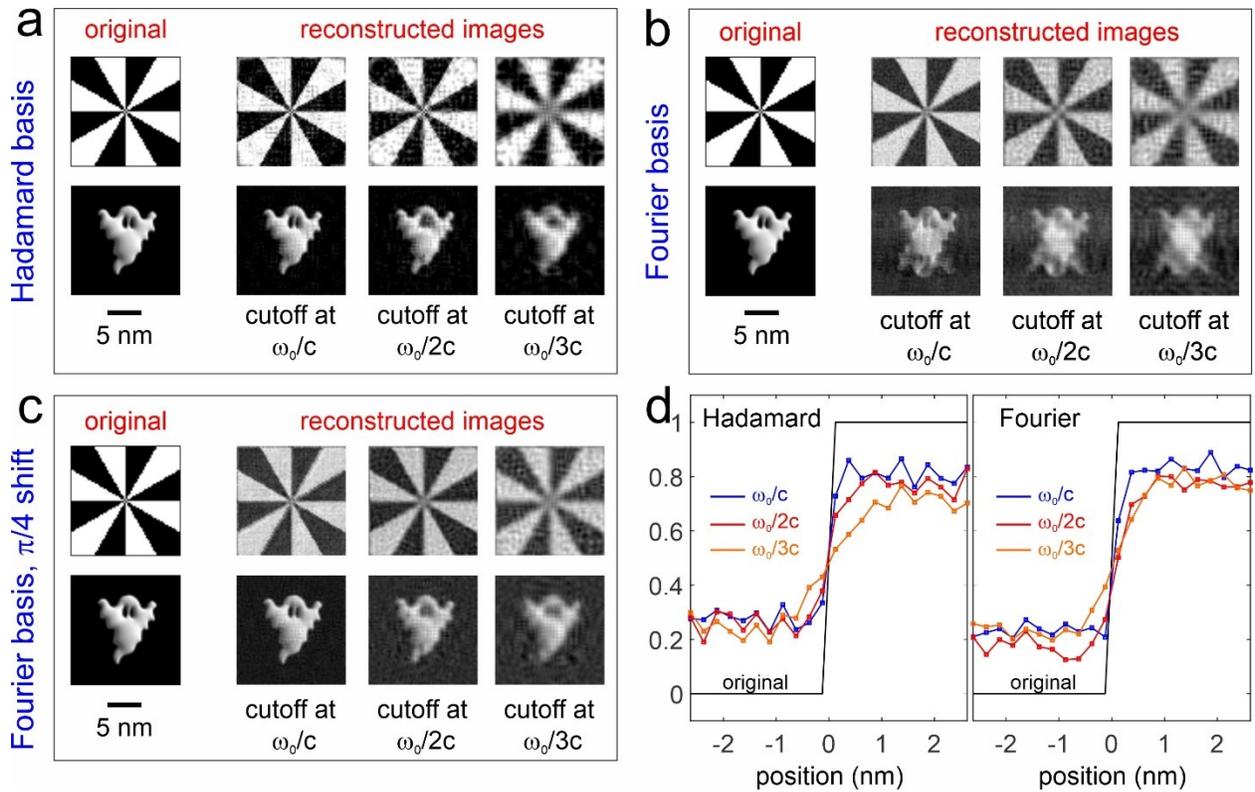

**Figure 3. ESPI imaging using Hadamard and Fourier bases.** Image reconstruction of a Siemens-star and a ghost-image performed using a Hadamard basis (*panel a*), a Fourier basis (*panel b*), and a Fourier basis with a π/4 phase shift (*panel c*). Reconstructed images are shown for different momentum cutoffs ($\omega_0/nc$, where $n = 1, 2, 3$) on the retrieved light field. The Field-of-View of all images if 16 x 16 nm². *Panel d*: spatial profiles obtained at the sharp edge of the Siemens star when using a Hadamard basis (left) and a Fourier basis (right). The black curve represents the original image, while the blue, red, and orange curves are associated with frequency cutoffs of $\omega_0/c$, $\omega_0/2c$, and $\omega_0/3c$, respectively. The spatial resolution is estimated by taking the 10:90 value of the error function fit for each curve. We obtain the following resolutions: 0.29 nm at a cutoff of $\omega_0/c$, 0.49 nm at a cutoff of $\omega_0/2c$, and 1.43 nm at a cutoff of $\omega_0/3c$ for the Hadamard-reconstructed images; 0.25 nm at a cutoff of $\omega_0/c$, 0.63 nm at a cutoff of $\omega_0/2c$, and 1.01 nm at a cutoff of $\omega_0/3c$ for the Fourier-reconstructed images.

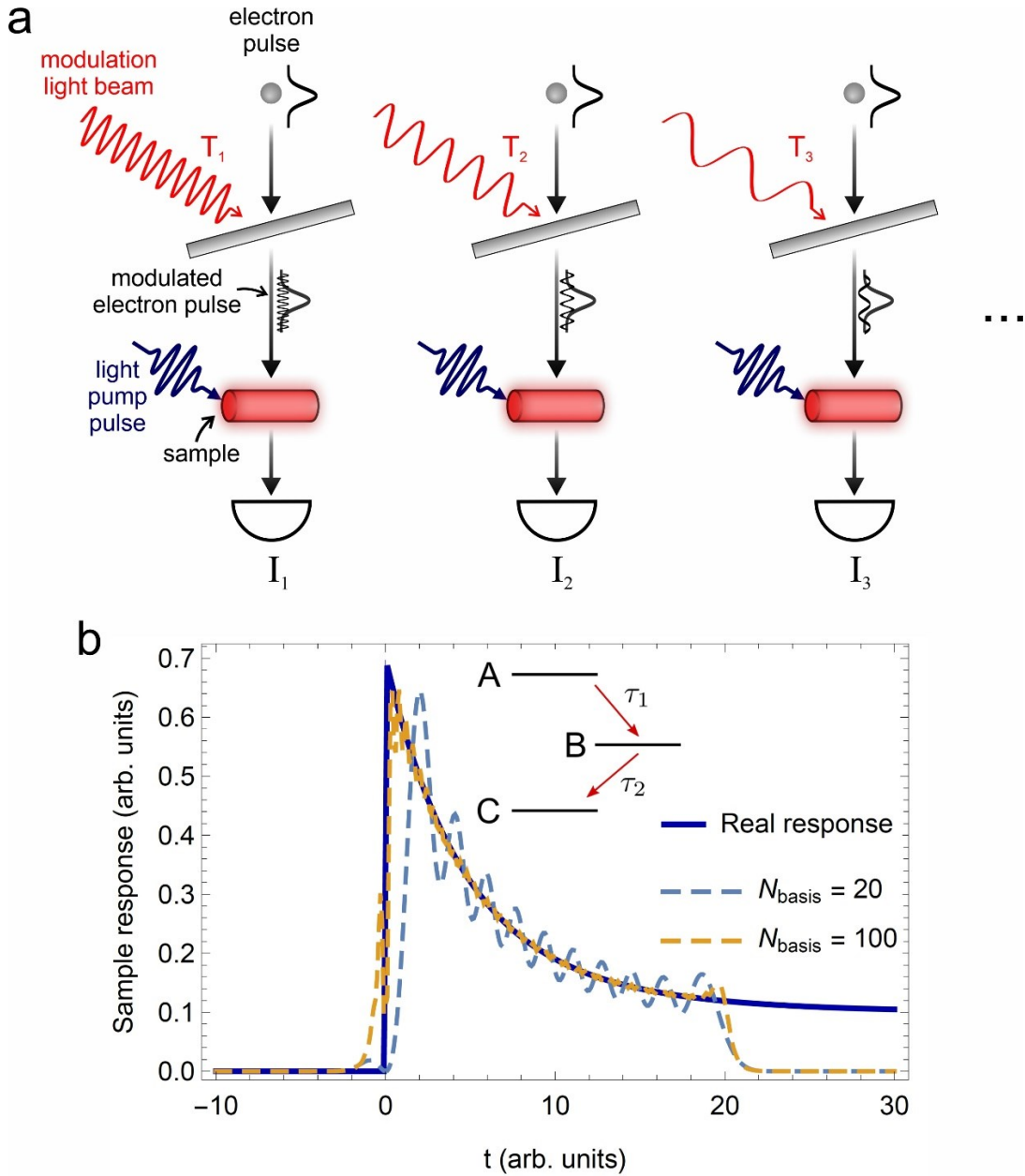

**Figure 4. Temporal Electron Single Pixel Imaging.** 1D temporal single-pixel reconstruction of a material dynamics. *Panel a*: A sequence of long light pulses with varying periods $T_j$ couple to the electron pulse via inverse transition radiation mediated by a metallic plate. We show three different periods: $T_1 < T_2 < T_3$. The longitudinally-modulated electron pulse then interacts with the sample in its excited state and, for each period $T_j$, a scattered intensity $I_j$ is measured. The full temporal evolution of the sample is finally reconstructed from a Fourier-like transformation of the measured signals (see main text for details). *Panel b*: simulated temporal dynamics of a system comprising three states (A, B, and C) according to the diagram in the inset. In the plot, the real response of the system (blue curve), obtained from a rate equation model, is compared with the results of temporal Fourier reconstructions using either 20 basis functions (dashed blue curve) or 100 basis function (dashed orange curve), as defined by Eq. (16).

# REFERENCES

[1] M. P. Edgar et al., Principles and prospects for single-pixel imaging, Nat. Photon. **13**, 13 (2019).

[2] M. F. Duarte et al. Single-pixel imaging via compressive sampling, IEEE Signal Process. Mag. **25**, 83–91 (2008).

[3] M. Gibson, S. D. Johnson, and M. J. Padgett, Single-pixel imaging 12 years on: a review, Opt. Express **28**, 28190-28208 (2020)

[4] C. A. Osorio Quero, D. Durini, J. Rangel-Magdaleno, and J. Martinez-Carranza, Single-pixel imaging: An overview of different methods to be used for 3D space reconstruction in harsh environments, Rev. Sci. Instrum. **92**, 111501 (2021).

[5] E. J. Candes, J. Romberg, and T. Tao, IEEE Trans. Inf. Theory **52**, 489 (2006).

[6] O. Katz, Y. Bromberg, and Y. Silberberg, Appl. Phys. Lett. **95**, 131110 (2009).

[7] L. Kovarik, A. Stevens, A. Liyu, and N. D. Browning, Implementing an accurate and rapid sparse sampling approach for low-dose atomic resolution STEM imaging, Appl. Phys. Lett. **109**, 164102 (2016).

[8] J. Schwartz, H. Zheng, M. Hanwell, Y. Jiang, R. Hovden, Dynamic compressed sensing for real-time tomographic reconstruction, Ultramicroscopy **219**, 113122 (2020).

[9] M.-J. Sun, L.-T. Meng, M. P. Edgar, M. J. Padgett & N. Radwell, A Russian Dolls ordering of the Hadamard basis for compressive single-pixel imaging, Scientific Reports **7**, 3464 (2017)

[10] W.-K. Yu, Super Sub-Nyquist Single-Pixel Imaging by Means of Cake-Cutting Hadamard Basis Sort, Sensors **19**, 4122 (2019)

[11] W.-K. Yu and Y.-M. Liu, Single-Pixel Imaging with Origami Pattern Construction, Sensors **19**, 5135 (2019)

[12] X. Yu, R. I. Stantchev, F. Yang & E. Pickwell-MacPherson, Super Sub-Nyquist Single-Pixel Imaging by Total Variation Ascending Ordering of the Hadamard Basis, Scientific Reports **10**, 9338 (2020)

[13] C. F. Higham, R. Murray-Smith, M. J. Padgett & M. P. Edgar, Deep learning for real-time single-pixel video, Scientific Reports **8**, 2369 (2018)

[14] Zibang Zhang, Xiang Li, Shujun Zheng, Manhong Yao, Guoan Zheng, and Jingang Zhong, Image-free classification of fast-moving objects using "learned" structured illumination and single-pixel detection, Optics Express 28, 13269-13278 (2020)

[15] B. W. Reed, A. A. Moghadam, R. S. Bloom, S. T. Park, A. M. Monterrosa, P. M. Price, C. M. Barr, S. A. Briggs, K. Hattar, J. T. McKeown, and D. J. Masiel, Electrostatic subframing and compressive-sensing video in transmission electron microscopy, Struct. Dyn. **6**, 054303 (2019).

[16] T Shimobaba, Y Endo, T Nishitsuji, T Takahashi, Y Nagahama, S Hasegawa, M Sano, R Hirayama, T Kakue, A Shiraki, T Ito, Computational ghost imaging using deep learning, Optics Communications **413**, 147-151 (2018)

[17] Wei Wang, Hao Wang, Haichao Wang, Guowei Li, Ni Chen & Guohai Situ, Deep-learning-based ghost imaging Meng Lyu, Scientific Reports **7**, 17865 (2017)

[18] Yuchen He, Gao Wang, Guoxiang Dong, Shitao Zhu, Hui Chen, Anxue Zhang & Zhuo Xu, Ghost Imaging Based on Deep Learning, Scientific Reports **8**, 6469 (2018)

[19] Fei Wang, Chenglong Wang, Chenjin Deng, Shensheng Han, and Guohai Situ, Single-pixel imaging using physics enhanced deep learning, Photonics Research **10**, 104-110 (2022)

[20] Shoupei Liu, Xiangfeng Meng, Yongkai Yin, Huazheng Wu, Wenjie Jiang, Computational ghost imaging based on an untrained neural network, Optics and Lasers in Engineering **147**, 106744 (2021)

[21] Yuchen He, Sihong Duan, Jianxing Li, Hui Chen, Huaibin Zheng, Jianbin Liu, Yu Zhou, Zhuo Xu, Ghost Imaging Based on Recurrent Neural Network, https://doi.org/10.48550/arXiv.2112.00736

[22] Fei Wang, Hao Wang, Haichao Wang, Guowei Li, and Guohai Situ, Learning from simulation: An end-to-end deep-learning approach for computational ghost imaging, Optics Express **27**, 25560-25572 (2019)

[23] Heng Wu, Ruizhou Wang, Genping Zhao, Huapan Xiao, Daodang Wang, Jian Liang, Xiaobo Tian, Lianglun Cheng, and Xianmin Zhang, Sub-Nyquist computational ghost imaging with deep learning, Optics Express **28**, 3846-3853 (2020)

[24] Q. Chen, C. Dwyer, G. Sheng, C. Zhu, X. Li, C. Zheng, Y. Zhu, Imaging Beam-Sensitive Materials by Electron Microscopy, Adv. Mater. **32**, 1907619 (2020).

[25] R. F. Egerton, Radiation damage to organic and inorganic specimens in the TEM, Micron **119**, 72-87 (2019).

[26] S. Li, F. Cropp, K. Kabra, T. J. Lane, G. Wetzstein, P. Musumeci, and D. Ratner, Electron Ghost Imaging, Phys. Rev. Lett. **121**, 114801 (2018)

[27] B. Barwick and D. J. Flannigan and A. H. Zewail, Photon-induced near-field electron microscopy, Nature **462**, 902-906 (2009).

[28] A. Feist, K. E. Echternkamp, J. Schauss, S. V. Yalunin, S. Schäfer & C. Ropers, Quantum coherent optical phase modulation in an ultrafast transmission electron microscope, Nature **521**, 200–203 (2015).

[29] G. M. Vanacore, I. Madan, F. Carbone, Spatio-temporal shaping of a free-electron wave function via coherent light–electron interaction, La Rivista del Nuovo Cimento **43**, 567-597 (2020)

[30] V. Di Giulio, M. Kociak, F. J. García de Abajo, Probing quantum optical excitations with fast electrons. Optica **6**, 1524−1534 (2019).

[31] O. Reinhardt, I. Kaminer, Theory of shaping electron wavepackets with light. ACS Photon. 7, 2859 (2020).

[32] K. E. Priebe, C. Rathje, S. V. Yalunin, T. Hohage, A. Feist, S. Schäfer, and C. Ropers, Attosecond electron pulse trains and quantum state reconstruction in ultrafast transmission electron microscopy, Nat. Photon. **11**, 793 (2017);

[33] M. Kozák, N. Schönenberger, and P. Hommelhoff, Ponderomotive generation and detection of attosecond free-electron pulse trains, Phys. Rev. Lett. **120**, 103203 (2018);

[34] Y. Morimoto and P. Baum, Diffraction and microscopy with attosecond electron pulse trains, Nat. Phys. **14**, 252 (2018)

[35] G. M. Vanacore et al., Attosecond coherent control of free-electron wave functions using semi-infinite light fields, Nat. Commun. **9**, 2694 (2018);

[36] M. Tsarev, A. Ryabov, P. Baum, Free-electron qubits and maximum-contrast attosecond pulses via temporal Talbot revivals. Phys. Rev. Res. **3**, 043033 (2021).

[37] C. Kealhofer, W. Schneider, D. Ehberger, A. Ryabov, F. Krausz, P. Baum, All-optical control and metrology of electron pulses, Science **352**, 429 (2016).

[38] G. M. Vanacore et al., Ultrafast generation and control of an electron vortex beam via chiral plasmonic near fields, Nat. Mater. **18**, 573–579 (2019)

[39] A. Feist, S. V. Yalunin, S. Schäfer, and C. Ropers, High-purity free-electron momentum states prepared by three-dimensional optical phase modulation, Phys. Rev. Research **2**, 043227 (2020).

[40] O. Schwartz, J. J. Axelrod, S. L. Campbell, C. Turnbaugh, R. M. Glaeser, H. Müller, Laser phase plate for transmission electron microscopy. Nat. Methods **16**, 1016−1020 (2019).

[41] I. Madan, G. M. Vanacore, S. Gargiulo, T. LaGrange, F. Carbone, The quantum future of microscopy: Wave function engineering of electrons, ions, and nuclei, App. Phys. Lett. **116**, 230502 (2020)

[42] A. Konečná, F. Iyikanat, F. J. García de Abajo, Entangling free electrons and optical excitations, arXiv:2202.00604

[43] A. Konečná and F. J. García de Abajo, Electron beam aberration correction using optical near fields, Physical Review Letters **125**, 030801 (2020).

[44] F. J. García de Abajo and A. Konečná, Optical modulation of electron beams in free space, Physical Review Letters **126**, 123901 (2021).

[45] I Madan, V Leccese, A Mazur, F Barantani, T LaGrange, A Sapozhnik, P M. Tengdin, S Gargiulo, E Rotunno, J-C Olaya, I Kaminer, V Grillo, F. J García de Abajo, F Carbone, and G M Vanacore, Ultrafast Transverse Modulation of Free Electrons by Interaction with Shaped Optical, ACS Photonics 2022, https://doi.org/10.1021/acsphotonics.2c00850

[46] M. C. C. Mihaila; P. Weber, M. Schneller, L. Grandits, S. Nimmrichter, T. Juffmann, Transverse Electron Beam Shaping with Light. Phys. Rev. X **12**, 031043 (2022).

[47] T. Sekia, Y. Ikuhara, N. Shibata, Theoretical framework of statistical noise in scanning transmission electron microscopy, Ultramicroscopy **193**, 118-125 (2018).

[48] N. Mevenkamp, P. Binev, W. Dahmen, P. M. Voyles, A. B. Yankovich and B Berkels, Poisson noise removal from high-resolution STEM images based on periodic block matching, Advanced Structural and Chemical Imaging **1**, 3 (2015)

[49] A. Kallepalli et al., Ghost imaging with electron microscopy inspired, non-orthogonal phase masks, DOI:10.21203/rs.3.rs-1111193/v1

[50] L Piazza et al., Simultaneous observation of the quantization and the interference pattern of a plasmonic near-field, Nat. Commun. **6**, 6407 (2015)

[51] I. Madan et al., Holographic imaging of electromagnetic fields via electron-light quantum interference, Sci. Adv. **5** eaav8358 (2019)

[52] Y. Morimoto and P. Baum, Attosecond control of electron beams at dielectric and absorbing membranes, Phys. Rev. A **97**, 033815 (2018).

[53] Xinyao Hu, Hao Zhang, Qian Zhao, Panpan Yu, Yinmei Li, and Lei Gong, Single-pixel phase imaging by Fourier spectrum sampling, Appl. Phys. Lett. **114**, 051102 (2019).

[54] Z. Zhang, X. Wang, G. Zheng, and J. Zhong, Hadamard single-pixel imaging versus Fourier single-pixel imaging, Optics Express **25**, 19619-19639 (2017).

[55] O. Kfir, H. Lourenço-Martins, G. Storeck, M. Sivis, T. R. Harvey, T. J. Kippenberg, A. Feist & C. Ropers, Controlling free electrons with optical whispering-gallery modes, Nature **582**, 46–49 (2020).

[56] K. Wang, R. Dahan, M. Shentcis, Y. Kauffmann, A. Ben- Hayun, O. Reinhardt, S. Tsesses, I. Kaminer, Coherent Interaction between Free Electrons and Cavity Photons, Nature **582**, 50 (2020).

[57] R. Dahan, S. Nehemia, M. Shentcis, O. Reinhardt, Y. Adiv, X. Shi, O. Be'er, M. H. Lynch, Y. Kurman, K. Wang and I. Kaminer, Resonant phase-matching between a light wave and a free-electron wavefunction, Nature Physics **16**, 1123-1131 (2020)

[58] R. Dahan, A. Gorlach, U. Haeusler, A. Karnieli, O. Eyal, P. Yousefi, M. Segev, A. Arie, G. Eisenstein, P. Hommelhoff, and I. Kaminer, Imprinting the quantum statistics of photons on free electrons, Science **373**, 1309-1310, (2021)

[59] J.-W. Henke, A. Sajid Raja, A. Feist, G. Huang, G. Arend, Y. Yang, F. J. Kappert, R. Ning Wang, M. Möller, J. Pan, J. Liu, O. Kfir, C. Ropers & T. J. Kippenberg, Integrated photonics enables continuous-beam electron phase modulation, Nature **600**, 653–658 (2021)

# Single-Pixel Imaging in Space and Time with Optically-Modulated Free Electrons

Andrea Konečná[1,2], Enzo Rotunno[3], Vincenzo Grillo[3], F. Javier García de Abajo[1,4,*], and Giovanni Maria Vanacore[5,*]

1. ICFO-Institut de Ciencies Fotoniques, The Barcelona Institute of Science and Technology 08860 Castelldefels (Barcelona), Spain
2. Central European Institute of Technology, Brno University of Technology, 612 00 Brno, Czech Republic
3. Centro S3, Istituto di Nanoscienze-CNR, 41125 Modena, Italy
4. ICREA-Institució Catalana de Recerca i Estudis Avançats, Passeig Lluís Companys 23, 08010 Barcelona, Spain
5. Laboratory of Ultrafast Microscopy for Nanoscale Dynamics (LUMiNaD), Department of Materials Science, University of Milano-Bicocca, Via Cozzi 55, 20121 Milano (Italy).

* To whom correspondence should be addressed:
javier.garciadeabajo@nanophotonics.es, giovanni.vanacore@unimib.it

**COUPLING COEFFICIENT BETA FOR A HOMOGENEOUS THIN FILM**

The coupling coefficient $\beta$ [see Eq. (2) in the main text] for a superposition of *p*-polarized light waves impinging from the $z < 0$ region with angle $\theta$ relative to the normal *z* direction on a self-standing homogeneous thin film of thickness $d$ can be written

$$\beta^m(\mathbf{R}) = \int_{-k_0}^{k_0} dk_x \int_{-k_0}^{k_0} dk_y \, exp(i\mathbf{K} \cdot \mathbf{R}) \beta_\mathbf{K}^m, \tag{S1}$$

which incorporates the fact that, due to the finite wavelength of light $\lambda = 2\pi/k_0 = 2\pi c/\omega$, we can imprint patterns with a precision limited by diffraction (i.e., the integral over transverse light wave vectors $\mathbf{K} = (k_x, k_y)$ is limited to the range $K < k_0$). The contributions of the different transverse wave vector components are

$$\begin{aligned}\beta_\mathbf{K}^m = \frac{ieK}{\hbar\omega k_0} \alpha_\mathbf{K} \Bigg[ &\frac{1}{\omega/v - k_z} + \frac{r_p}{\omega/v + k_z} - \frac{t_p \exp(-i\omega d/v)}{\omega/v - k_z} \\ &+ A\frac{\exp[i(-\omega/v + k_z')d] - 1}{\omega/v - k_z'} + B\frac{\exp[-i(\omega/v + k_z')d] - 1}{\omega/v + k_z'} \Bigg].\end{aligned} \tag{S2}$$

Here, the coefficients $\alpha_\mathbf{K}$ are controlled through the proper settings of the spatial light modulator, and $r_p$, $t_p$, $A$, and $B$ are given by

$$r_p = r_p^0 \left[ 1 - \frac{(t_p^0)^2 (k_z'/k_z) \exp(2ik_z'd)}{1 - (r_p^0)^2 \exp(2ik_z'd)} \right], \tag{S3}$$

$$t_p = \frac{(t_p^0)^2 (k_z'/k_z) \exp(ik_z'd)}{1 - (r_p^0)^2 \exp(2ik_z'd)}, \tag{S4}$$

$$A = \frac{1}{\sqrt{\epsilon}} \frac{t_p^0}{1 - (r_p^0)^2 \exp(2ik_z'd)}, \tag{S5}$$

$$B = \frac{1}{\sqrt{\epsilon}} \frac{-t_p^0 r_p^0 \exp(2ik_z'd)}{1 - (r_p^0)^2 \exp(2ik_z'd)}, \tag{S6}$$

where $r_p^0$ and $t_p^0$ are the Fresnel reflection coefficients expressed in terms of the permittivity of the film material $\epsilon$ as

$$r_p^0 = \frac{\epsilon k_z - k_z'}{\epsilon k_z + k_z'}, \tag{S7}$$

$$t_p^0 = \frac{2\sqrt{\epsilon} k_z}{\epsilon k_z + k_z'}, \tag{S8}$$

with out-of-plane light wave vector components $k_z$ and $k_z'$ given by

$$k_z = \sqrt{k_0^2 - K^2}, \tag{S9}$$

$$k_z' = k_0 \sqrt{\epsilon - K^2/k_0^2}. \tag{S10}$$

In the perfect-electric-conductor (PEC) limit for the film material, we have $r_p = r_p^0 = 1$ and $t_p = 0$, so the **K**-dependent amplitudes of the coupling coefficient reduce to

$$\beta_{\mathbf{K},\text{PEC}}^m = \frac{2ieK}{\hbar k_0 v} \frac{\alpha_\mathbf{K}}{(\omega/v)^2 - (k_z)^2}. \tag{S11}$$

## TEMPORAL ELECTRON SINGLE-PIXEL IMAGING

We consider a system comprising three states ($j$ = A, B, and C) according to the diagram in Fig. 4b of the main text. At time zero, the system is taken to be pumped to an excited state A, from which it decays in a cascade fashion to B and then to C. The time evolution of the populations of the three states within our model system, $p_A, p_B$ and $p_C$ (shown in the inset of Fig. 4d), is governed by the rate equations

$$\dot{p}_A = -\frac{p_A}{\tau_1}, \qquad \dot{p}_B = \frac{p_A}{\tau_1} - \frac{p_B}{\tau_2}, \qquad \dot{p}_C = \frac{p_B}{\tau_2}, \tag{S12}$$

which are supplemented by the initial conditions $p_A(0) = 1$, $p_B(0) = p_C(0) = 0$. Here, $\tau_1$ and $\tau_2$ are the lifetimes associated with the decay from A to B and from B to C, respectively. The solution to this set of equations is

$$p_A = e^{-\frac{t}{\tau_1}},$$

$$p_B = \frac{1}{\frac{\tau_1}{\tau_2} - 1}\left(e^{-\frac{t}{\tau_1}} - e^{-\frac{t}{\tau_2}}\right), \tag{S12}$$

$$p_C = \frac{1}{\frac{\tau_1}{\tau_2} - 1}\left(-\frac{\tau_1}{\tau_2}e^{-\frac{t}{\tau_1}} + e^{-\frac{t}{\tau_2}}\right) + 1.$$

We then consider the time evolution of the measured scattering intensity $I$ for electrons interacting with the system at time $t$:

$$I(t) = \Theta(t) \sum_{j \in \{A,B,C\}} p_j(t) a_j =$$

$$= \Theta(t) \left\{ a_C + e^{-\frac{t}{\tau_1}} \left[ a_A + \frac{1}{\frac{\tau_1}{\tau_2} - 1}\left(a_B - \frac{\tau_1}{\tau_2} a_C\right) \right] + e^{-\frac{t}{\tau_2}} \frac{a_C - a_B}{\frac{\tau_1}{\tau_2} - 1} \right\}, \tag{S14}$$

where $\Theta(t)$ is the Heaviside function and the constants $a_j$ are the intensities observed when the system is in state $j$. This expression implicitly assumes that the interaction time per electron is short compared with the decay times.

We use Fourier-like basis functions for the evolution of the incident electron current as a function of time with respect to the pumping time:

$$H^m(t) = e^{-\frac{\left(t - \frac{t_{\max} + t_{\min}}{2}\right)^2}{2\sigma^2}} \sin^2[\pi m \, t/(t_{\max} - t_{\min})], \tag{S15}$$

where $t_{\min}$ and $t_{\max}$ determine the boundaries of the sampling time interval and $\sigma^2$ is the variance of the envelope of the probing electron wave function. In our example, we set $\sigma = 0.1(t_{\max} - t_{\min})$, $a_A = 0.7, a_B = 0.2, a_C = 0.1$, $\tau_1 = 4$, and $\tau_2 = 8$, all in arbitrary units. The reconstruction is then performed in an analogous way to the spatial domain. We again consider the non-orthogonality of the illumination basis.

In Fig. 4b, we show the reconstructed time profile in the defined interval using either 20 or 100 basis functions, which are able to correctly retrieve the real response of the system. From an experimental viewpoint, this approach can be used to retrieve the transient dynamics of a material in the few-femtosecond range even when using very long electron pulses (picosecond or longer).

## NOTES ON OPTIMAL DISCRIMINATION AND USE OF *A PRIORI* INFORMATION IN SPI AND CONVENTIONAL IMAGING

We provide here a more specific example of how *a priori* information can benefit ESPI, especially in terms of optimal discrimination, but less so for post-processing in conventional raster-scanning imaging. For instance, the knowledge that a given sample is sparse in real space (e.g., a few particles dispersed in a large area) is already used in ESPI, in which each measurement of a projection on a given base function gives more information than in conventional imaging, where many measurements of empty pixels are needed before some nonzero signal is detected.

In general, it is non-trivial to make a direct comparison between the two techniques. Conventional raster scanning has the advantage of its simplicity and direct interpretation. However, we identify here a specific example that illustrates some conditions under which ESPI holds a direct advantage.

The key is the use *a priori* information. Raster scanning is assumed to be done with a strategy that is object-independent, while SPI allows one to optimize the acquisition strategy even before starting the experiment. In raster scanning any *a priori* information is applied only after acquisition to interpret the image, something that for conventional imaging can be regarding as de-noising.

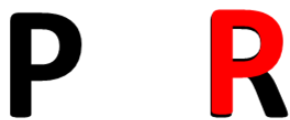

In this comparison we push the condition to the extreme by considering a very strong *a priori* information, and thus, we are left with only two possibilities. Let's consider that our amplitude sample looks like a letter 'P' or a letter 'R', as in the figure above. The difference between the two is in the right-hand leg of the 'R', while the rest of the structure is identical. Standard raster-scanning must necessarily cover also the 'P', as it does not use, by definition, any *a priori* information before performing the experiment. Raster scanning is therefore intrinsically less efficient, and no post-acquisition strategy can remedy the loss of counts in the 'P' body (i.e., a region that does not add anything to the known information). While ESPI could solve the ambiguity with a single electron, raster-scanning could easily be non-conclusive even when using a few of them.

In a recent paper by Troiani *et al*. [Phys. Rev. A **102**, 043510 (2020)], the authors have highlighted that, in order to optimize an electron measurement, the detection must concentrate on the projection on the differences between the two states. In this simplified case, the optimization would thus just concentrate on the shape difference because intensity levels are binary. In the P-R discrimination, the difference is just the right-hand "leg" of the R and it is a 0 or 1 discrimination.

SPI would thus use this *a priori* information to reduce the amount of patterns and electrons used in the reconstruction, whereas conventional imaging would still scan over the entire image and use such *a priori* information only for a post-acquisition de-noising.

To convey a more general argument, we can also consider objects with different levels of transparency (and not only 0 and 1 as in the previous example), where an optimization strategy would be more complex. Let's assume, for instance, that the sample is composed of two parts and, as before, our *a priori* information is very strong, so we are left with only two possible samples to discriminate. For the sample S1 (see figure below), the transmittance is 0.75 in the left part and 0.25 in the right part. The second sample S2 is instead the opposite: 0.25 transmittance on the left and 0.75 transmittance on the right. Once again both samples are only amplitude objects.

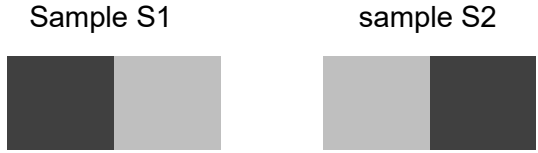

The optimal measurement theory (see, for instance, Troiani et al., Phys. Rev. A **102**, 043510 (2020)) indicates that the ideal measurement relies on the projection (scalar product) on a vector that would produce a difference between the two objects (i.e., the vector **v** = (1,-1) in this case). We assume here that the result of two measurements in the two parts can be summed together (and that the phase of the wave addressing the two parts is controllable). This is experimentally possible only for a SPI scheme in which the detector is in a diffraction plane.

For a experiment, we assume that we can only measure the intensity or amplitude of such projection. Then, the absolute value of the projection of the sample S1 on the vector **v** would give |0.75-0.25| = 0.5, and for the sample S2 this projection would also be |0.25-0.75| = 0.5. This is the same value in the two cases. The optimal discrimination theory thus tells us that the best projector vector can be constructed as **v'** = **v** + (0.5,0.5), and therefore, the two projections for the samples S1 and S2 on the vector **v'** will now be 1 and 0, respectively. Namely, using this projector we would have a "1 or 0" type of experiment even if the transmissivity is not binary in real space. A fundamental reason for this is that SPI is compatible with an optimization in a multiplicity of bases, while raster scanning can only be used to optimize in real space.

**THREE-STEP ALGORITHM WITH FOURIER BASIS**

When using a Fourier basis, the reconstruction can be alternatively performed with a three-step (or alternatively four-step) algorithm (see Ref. 33 in the main text), where we define

$$\Delta_\varphi(\mathbf{K}) = \int d^2\mathbf{R_S} T(\mathbf{R_S}) H(\mathbf{R_S}, \mathbf{K}, \varphi). \tag{S16}$$

Then, we calculate

$$\Delta_{\text{Tot}} = \left(2\Delta_0 - \Delta_{2\pi/3} - \Delta_{4\pi/3}\right) + i\sqrt{3}\left(\Delta_{2\pi/3} - \Delta_{4\pi/3}\right) \tag{S17}$$

and finally perform the inverse Fourier transform to reconstruct the sample transmission function

$$T(\mathbf{R_S}) = \text{FT}^{-1}\{\Delta_{\text{Tot}}(\mathbf{K})\}, \tag{S18}$$

which can be done after collecting the intensities at the detector obtained with varying spatial frequencies.

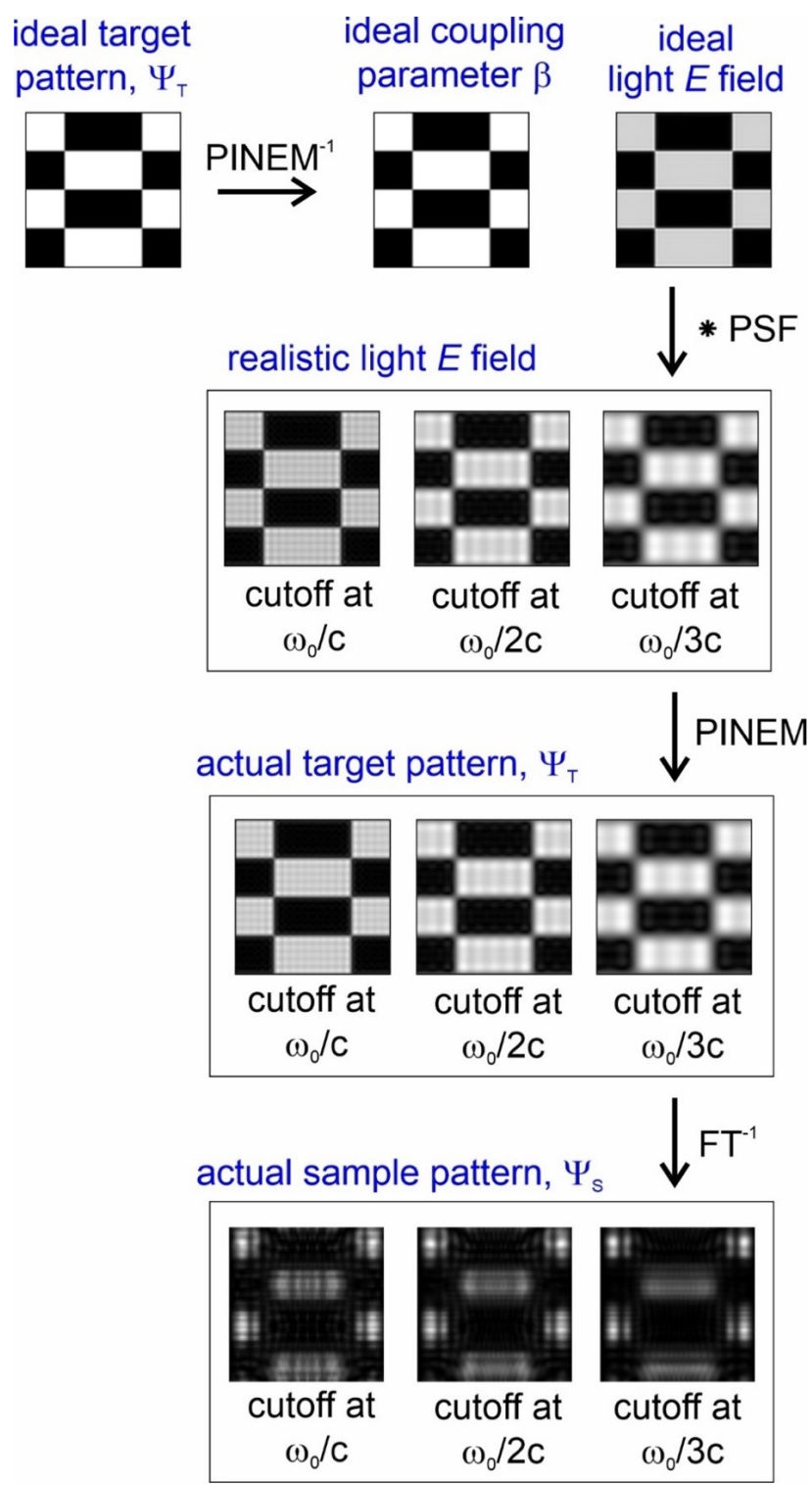

**Figure S1.** Sequence of operations used to calculate the transverse distribution of beam electrons arriving on the sample when starting from an ideal target pattern and considering realistic non-ideal conditions. Here, we plot results for a pattern taken from a Hadamard basis.

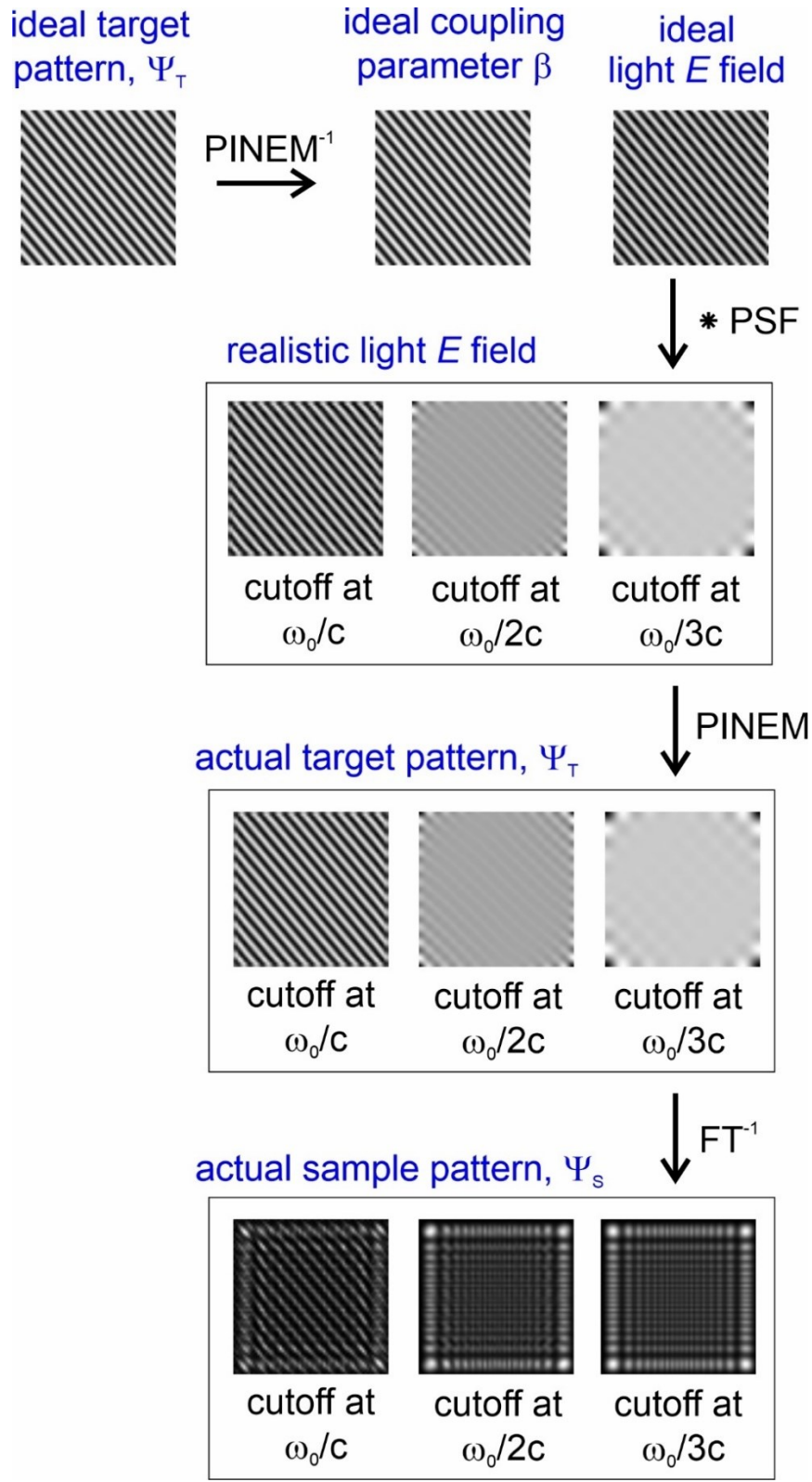

**Figure S2.** Sequence of operations used to calculate the transverse distribution of beam electrons arriving on the sample when starting from an ideal target pattern and considering realistic non-ideal conditions. Here, we plot results for a pattern taken from a Fourier basis.

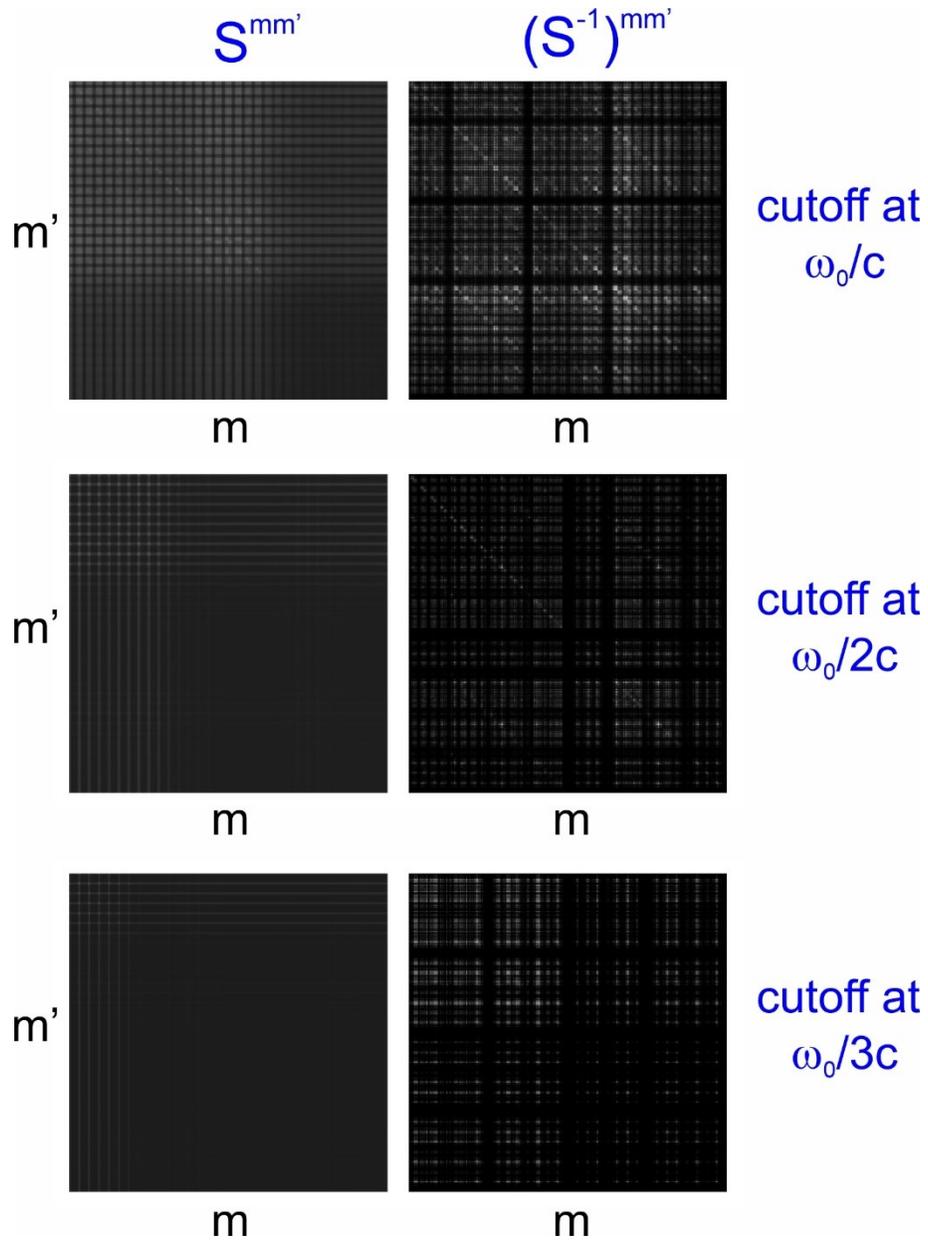

**Figure S3.** Overlap matrix calculated for the Hadamard basis when considering imperfect illumination of the sample plotted for the three different cutoffs considered in this work.